# The effect of configurational complexity in heteropolymers on the coil-globule phase-transition

Fabrizio Tafuri[1], Andrea M. Chiariello[1,*]

[1]Dipartimento di Fisica, Università di Napoli Federico II, and INFN Napoli, Complesso Universitario di Monte Sant'Angelo, 80126 Naples, Italy

*Correspondence: Andrea M. Chiariello. E-mail address: chiariello@na.infn.it

**Abstract**. The coil-globule transition of hetero-polymer chains is studied here. By means of extensive Molecular Dynamics simulations, we show that the transition is directly linked to the complexity of the chain, which depends on the number of chemical species defined in the environment and the location of the binding sites along the polymer. In addition, when the number of species increases, we find that the distribution of binding sites plays an important role in triggering the transition, beyond the standard control parameters of the polymer model, i.e. binders concentration and binding affinity. Overall, our results show that by increasing the system complexity new organizational layers emerge, thus allowing a more structured control on the polymer thermodynamic state. This can be potentially applied to the study of chromatin architecture, as such polymer models have been broadly used to understand the molecular mechanisms of genome folding.

## INTRODUCTION

Understanding how genome folds in the nucleus of cells is one of the most important questions in Molecular Biology. Indeed, the development of powerful experimental technologies ([1]) allowed to show that chromosomes exhibit a complex architecture, occurring at different genomic length scales in the cell nucleus ([2]-[5]). Specifically, chromatin is folded into Mb-sized domains called topologically associating domains (TADs, [6]-[7]); TADs have internal structure (sub-TADs) and are included in larger domains, named A and B compartments ([8]), associated with open and active chromatin and more compact and inactive chromatin, respectively. Also, TADs form a non-random hierarchy known as metaTAD structure ([9]). Such spatial structure of chromatin is linked to cellular activities. For example, distant regulatory elements (enhancers) located along DNA sequences trigger the functions of target genes through physical contact. Anyway, the mechanisms governing such complex and precise organization are not fully understood.

To tackle the organization problem at different length scales, polymer models have been successfully employed. Indeed, quantities of interest from Polymer Physics prove useful in understanding the complexity of spatial organization. For example, by means of microscopy-based experiments (e.g. by FISH or STORM methods, [10]) the spatial distances between several loci can be simultaneously measured and compared with model predictions ([11]-[12]). In addition, the Hi-C technique ([8]) allows to measure the frequency with which two DNA fragments physically co-localize in space. Again, from the Hi-C contact matrix we can calculate quantities (e.g. the mean contact probability of two regions as a function of their linear distance) that can be compared with polymer model and reveal information about the physical properties of the chromatin in a specific genomic region. Here we focus on the String and Binders Switch (SBS) polymer model ([13]), which have been successfully employed to describe several features of genome folding, as contact probability scaling ([14]), spatial organization of real loci ([15]) or multiple contacts ([11]). In the SBS scenario, the contacts between distant regions are mediated by diffusive molecules (named binders) that can attractively interact with binding sites located along the polymer.

The position and the number of types of binding sites define the complexity of the polymer and can be conveniently set to describe different contexts, as simple homo-polymers (a uniform polymer with just one type of binding sites), block co-polymers (two or more types of binding types arranged in linear segments along the polymer) or hetero-polymers with multiple binding sites located along the polymer. In this regard, computational methods ([16]-[18]) have been designed to customize polymer models aiming to accurately describe genome architecture. Using machine learning approaches fed with experimental data, e.g. Hi-C contact matrices, it is possible to infer a polymer model that best reproduces the degree of compaction of a specific genomic region of interest, i.e. the minimum number of different types of binding sites and their spatial distribution along the hetero-polymer chain ([19]).

In this work, we use massive parallel molecular dynamics simulations to explore, in a highly controlled polymer system, the role of configurational degeneracy of hetero-polymers in the coil-globule phase transition. We start from a simple, uniform homo-polymer and gradually increase the complexity of the chain by introducing different types of binding sites. Then, we study the physical properties of the system in each condition at equilibrium, focusing on the thermodynamic state of the polymer (coil or globule). We find that the folding state of an alternating hetero-polymer is affected by the number of types of binding sites, for fixed binding affinity and binder concentration. Indeed, more complex polymers require higher interaction affinities to fold into the globular state. Furthermore, we find that complex polymers exhibit different equilibrium states when different arrangements of the binding sites are considered. Hetero-polymers with random arrangements of binding sites can transit into the globular state, whereas hetero-polymers with regular alternating sequences of different types do not, being the other control parameters unchanged. This implies that the folding state of complex polymers may be controlled, beyond the binder concentration and binding affinity, through the configuration of the binding sites along the polymer chain. In a biological framework, different binding types are typically mapped into different molecular signatures. Our study suggests that their abundance found in cell nuclei could be another regulatory mechanism to control genome folding and gene activity.

**RESULTS**

**The system.** To investigate the general properties of the coil-globule phase transition in hetero-polymers, we consider a polymer model equipped with binding sites able to attractively interact with diffusive molecules (binders) which populate the surrounding environment. Biologically, the polymer represents a chromatin filament, the binding sites represent genomic regulatory elements, as enhancer and promoters, or binding sites associated to protein binding motifs and the binders represent the molecular factors that constantly bind to the chromatin in the cell nucleus (such as transcription factors) ([20]). In a very simple version of this model (generally known as the SBS model, [13]), the polymer consists of one binding sites type and inert, non-interacting beads (schematically represented as red and grey bead respectively in **Fig. 1**, **a**). Nevertheless, as several different proteins exist in cell nuclei, the model can be naturally generalized to envisage the existence of more binding site types, associated to an equal number of cognate types of binders. In any case, the system control parameters are the binder concentration c and binding affinity $E_{int}$. If both c and $E_{int}$ are above threshold, the system undergoes the coil-globule phase transition ([13], [14]) and the polymer collapses in a stable, globular phase. In the following, we will explore such thermodynamic behaviour by varying the polymer complexity, given by the number of types of binding sites (schematically, the number of colors) and their spatial distribution along the sequence. Details about parameters of the simulations can be found in the Methods section.

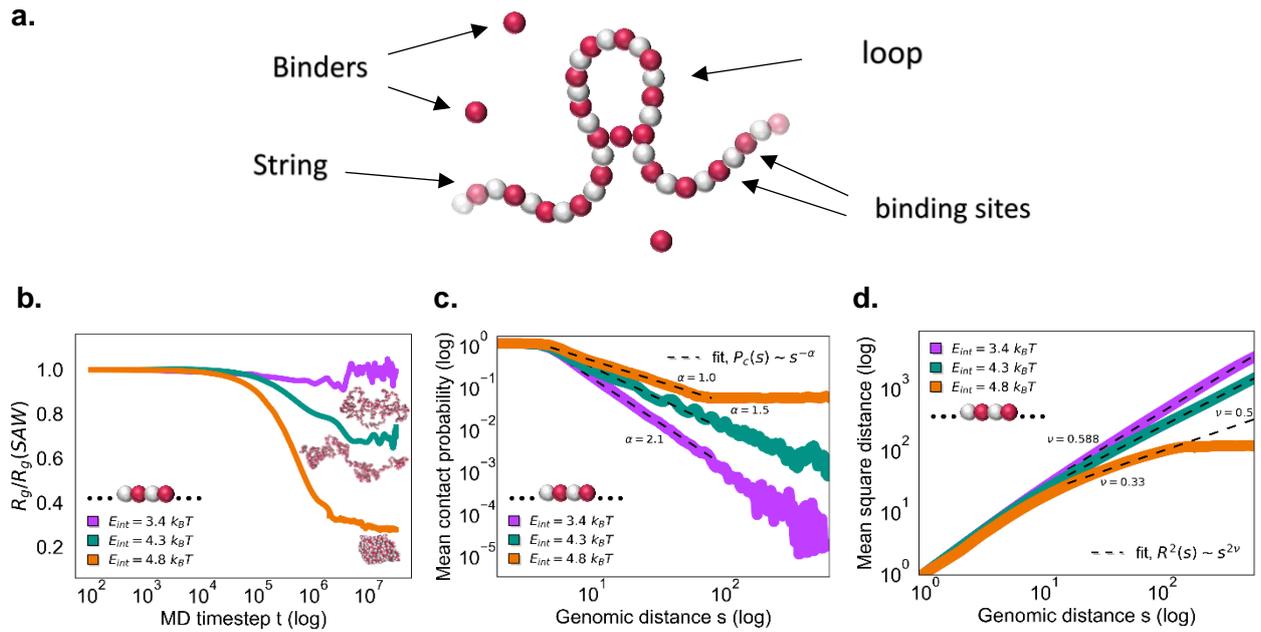

**Fig. 1 The coil-globule phase transition of a hetero-polymer SBS models. a.** A schematic cartoon of an SBS hetero-polymer model is shown. In this case, one type of binding site, colored in red, is present. **b, c, d.** Thermodynamic phases of an alternating SBS polymer with one type of binding site, as function of the binding affinity $E_{int}$. **b.** The time dynamics of normalized gyration radius shows that the transition occurs as the binding affinity increases. Examples of 3D snapshots taken from MD simulations are reported. **c.** Scaling of mean contact probability $P_c(s) \sim s^{-\alpha}$ in the three different thermodynamic equilibrium phases. **d.** Scaling of the mean square distance $R^2(s) \sim s^{2\nu}$. In the open SAW conformation ($E_{int} = 3.4\ k_BT$) the scaling exponents result $\alpha = 2.1$, $\nu = 0.588$, in the compact globule state ($E_{int} = 4.8\ k_BT$) $\alpha = 1.0$, $\nu = 0.33$ and in Θ-point thermodynamic phase ($E_{int} = 4.3\ k_BT$), $\alpha = 1.5$, $\nu = 0.5$. Binder concentration is fixed for all cases.

**Phase transitions of alternating hetero-polymer chains.** We first study the phases of hetero-polymers as a function of the number types of binding sites, with fixed binder concentrations and binding affinities. For this purpose, we simulate the time dynamics of alternating hetero-polymer chains, where an ordered subsequence of differently colored beads is repeated along the polymer (Methods). Additional effects of binding sites spatial distributions along the polymer will be considered in the following sections. The simplest model of alternating SBS hetero-polymer is the uniform homo-polymer, where the noninteracting bead (grey) and binding site (red) alternate along the string (**Fig. 1**, **a**). We performed simulated the MD dynamics of such system by varying the binding affinity, for fixed binder concentration. The goal was to identify the affinity $E_{int}$ values for which the homo-polymer chain can undergo the coil-globule transition. As usual, we used the gyration radius ([14], Methods) to monitor the transition during time. Indeed, when the transition occurs, the polymer size is reduced, and the gyration radius sharply drops (**Fig. 1**, **b**). In agreement with the predictions of Polymer Physics ([21], [22]), three thermodynamic states are found, with the transition threshold (Θ-point) identified around $E_{int}=4.3k_BT$. Consistently, in each phase the scaling exponents for mean contact probability and mean square distance (Methods) is compatible with the theoretical value (**Fig. 1**, **c**, **d**). Then, we gradually increase the complexity of the chains by adding more types (colors) of binding sites to the sequence. As first step, we use an alternating chain in which there are two types of binding sites (red and green) and use the same binder concentration for each type. To compare the behaviour at equilibrium of different hetero-polymer chains in the same condition, we set the binding affinity $E_{int}$ = 4.8 $k_B T$ which gave the transition for the chain with a single type of binding site. As before, we observe that with this affinity the chain with two types can transit into the globular phase (**Fig. 2**, **a**, green curve), as confirmed by the contact probability and mean square distance (**Fig. 2**, **b** and **c**, green curve). By repeating the previously process, we increase again the complexity and consider a chain with three binding site types (red, green and blue), following the same alternating scheme, at the same concentration and affinity $E_{int}$. Unlike the previously considered hetero-polymer chains, in this case the transition does not occur, as no substantial change in the gyration radius is detected from the initial SAW configuration (**Fig. 2**, **a**, cyan curve). Analogously, the equilibrium $P_c(s)$ and $R^2(s)$ clearly show that the polymer stays in the thermodynamic SAW phase (**Fig. 2**, **b** and **c**, cyan curves). Overall, our results show that the thermodynamic equilibrium phases of alternating hetero-polymer chains are importantly affected by the number of types of binding site, that is by its complexity. At the binding affinity $E_{int}$=4.8$k_B T$, the alternating hetero-polymer chain undergoes a phase transition to the compact state for only one and two types of binding site; the addition of a third (or obviously more) colors keeps the polymer in the open conformation and higher interaction affinities are therefore required to trigger the transition.

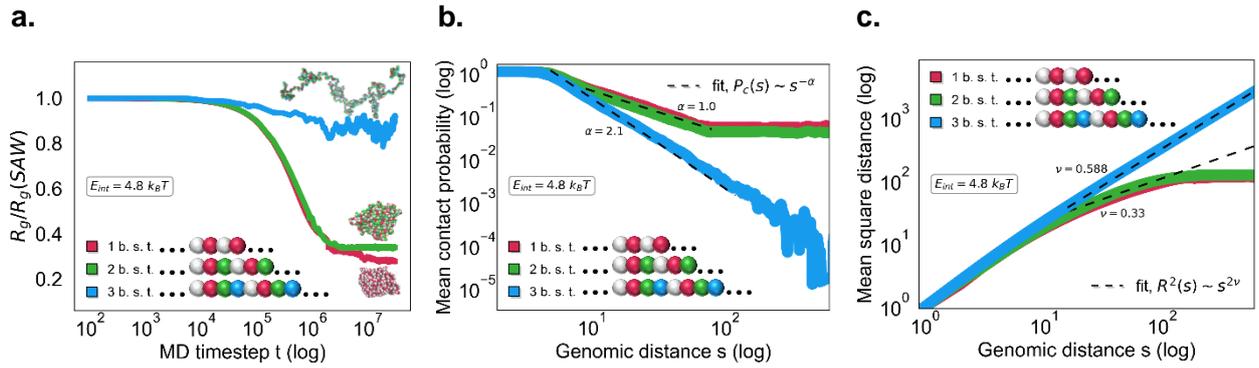

**Fig. 2 Number of types of binding sites is important in the phase transition.** The folding state of an alternating hetero-polymer chain depends on the number of binding site types (b.s.t) along the polymer. **a.** $R_g/R_g(SAW)$ dynamics is very similar for the 1 and 2 b.s.t. hetero-polymers (red and green curves) where the transition occurs. Conversely, in the 3 b.s.t alternating polymer (cyan curve) the transition does not occur. Examples of 3D equilibrium structures are also shown. **b.** and **c.** Equilibrium mean contact probability and mean square distance in the different hetero-polymers. Binding affinity ($E_{int}$ = 4.8 $k_B T$) and concentration is the same in all cases.

**Phase transitions of random hetero-polymer chains.** We have shown that the number of types along the chain plays an important role in triggering the phase transition. Next, we investigated the potential role in the arrangement of the binding sites in driving the phase transition. To this aim, we focused on hetero-polymer chains with three binding site types as before and rearranged the positions of its beads. As first rearrangement, we first considered random permutations of the binding sites, where the bead positions (including the grey ones) are randomly bootstrapped along the string (**Fig. 3**, **a**). MD simulations were then performed by keeping binding affinity $E_{int}$ = 4.8 $k_B T$ and binder concentration unchanged with respect the previous case. Interestingly, the results of the simulations revealed that the re-arranged polymer collapses into a stable globular shape, although the transition was not observed for the regular alternating chain. To check the robustness of the results, we considered three independent random permutations, and the transition was observed in all cases (**Fig. 3**, **b**, Methods). Next, we verified that such effect is not strictly dependent on the specific binding affinity. Indeed, simulations with lower binding affinity ($E_{int}$ = 4.1 $k_B T$) exhibit a similar behaviour, although the transition is not complete, and the polymer results compacted only locally (**Fig. 3**, **c**, Methods). These results point toward a scenario in which the configuration of the binding sites along the chain is another possible, general control mechanism to trigger the phase-transition (see next section) in this kind of systems. Again, such results have interesting implications for the chromatin architecture, as the regulation of some genomic regions can be orchestrated not only by the mere presence of the regulatory elements, but it is also encoded in their specific position along the genome. The complexity of the polymer therefore allows a more precise control on gene activity. In the next section, we investigate the microscopic origin of this process and give a mechanistic insight to it.

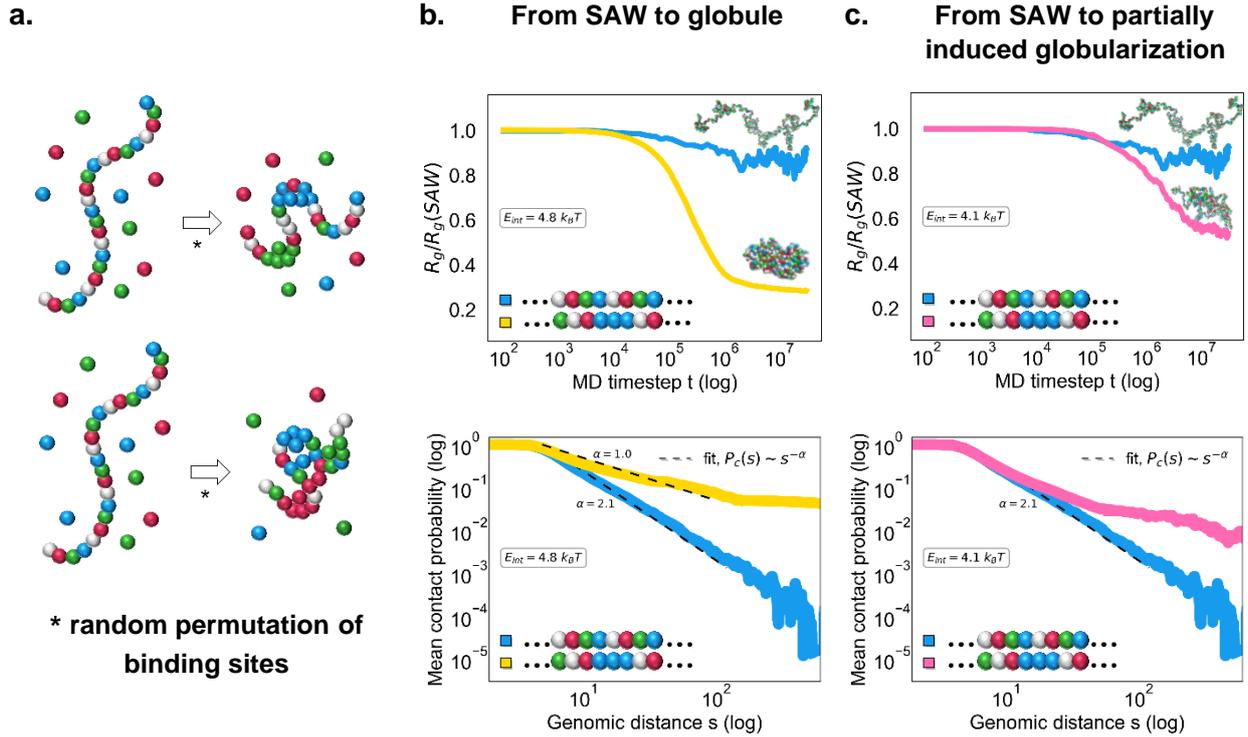

**Fig. 3 Rearrangement of binding site positions induces phase transition. a.** Schematic cartoons of alternating and randomly permuted SBS polymers. Alternating hetero-polymers do not undergo the coil-globule transition (left), whereas randomly rearranged hetero-polymers can do (right). **b.** Normalized gyration radius dynamics (top) and equilibrium contact probability (bottom) of alternating (cyan curve) and random hetero-polymers (yellow curve) at binding affinity $E_{int}$=4.8 $k_B$T. Examples of 3D snapshots taken from MD simulation at equilibrium are reported. **c.** as in b, for $E_{int}$=4.1 $k_B$T. Curves are an average over three different independent random permutations.

**Microscopic origins of the phase transition.** We then asked about the microscopic mechanism which originate such a different behaviour observed in the alternating and the random permuted polymer, although they carry the same number of binding sites for each type. We hypothesized that in the permuted configurations there exist regions with consecutive beads of the same type (not present in the alternating polymer) which can act as starting point to trigger the phase transition. Therefore, to quantitatively evaluate this configurational, we considered the following strategy: in brief, we start from alternating hetero-polymer chains gradually created pairs of consecutive binding sites, by swapping the position of selected beads (**Fig. 4**, **a**, see Method for the details of the developed algorithm). Then, we studied how the polymer folding state varies as a function of the number of pairs of binding sites of the same (any) color, at a fixed binding affinity $E_{int}$ and concentration. Since in the previous section we studied alternating and random chains composed of three binding site types and an inert (grey) species, this strategy allowed to consider intermediate, highly controlled configurations. Interestingly, we find that the polymer starts to be able to transit into the globular state after a threshold number of pairs (estimated around 100, i.e. 50 bead swaps) are formed, as shown in **Fig. 4 b** (here $E_{int}$=4.8$k_B$T). It is also worth to note that in a such condition a few pairing events (fraction of 0.1-0.2, see Methods) are required to trigger the transition (see next section), highlighting that macroscopic effects can be observed from microscopic changes of the system ([23]). Overall, these results suggest that the presence of regions with consecutive beads of the same type enhance at the microscopic scale the avidity of the polymer to attract binders, which in turn can form more stable loops and drive the polymer in the globular state.

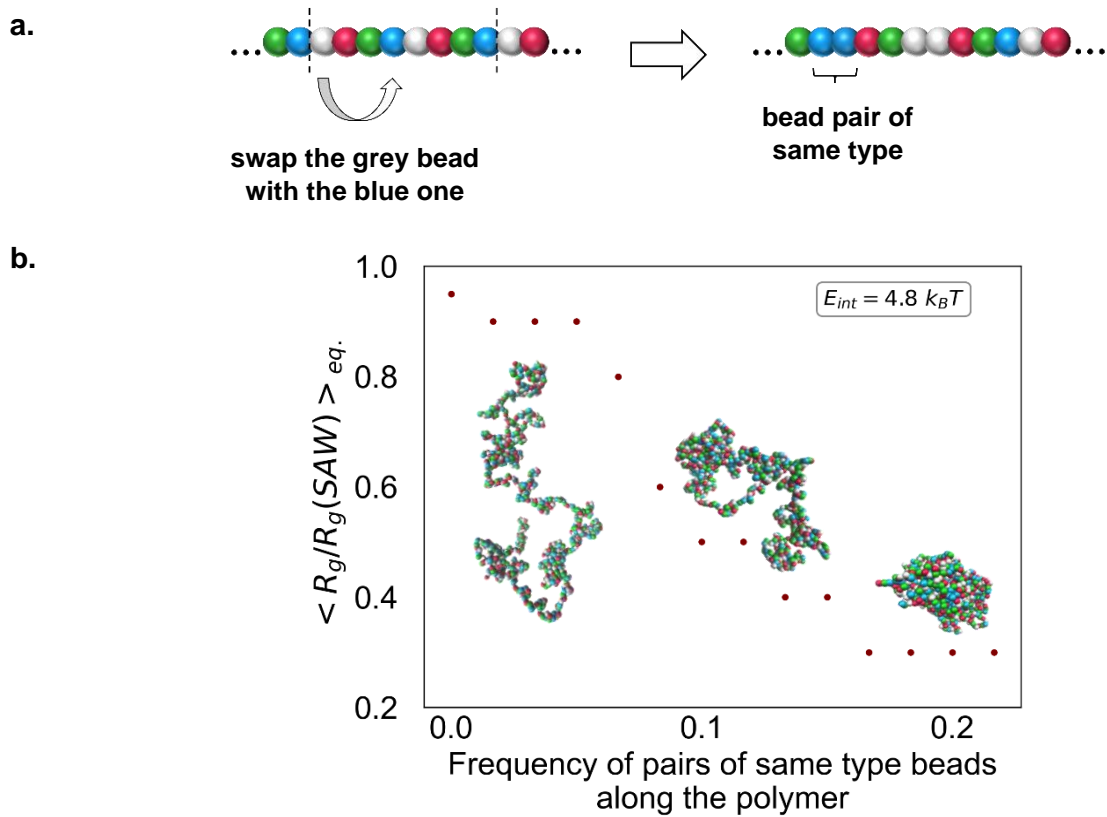

**Fig. 4 Frequency of homolog pairs triggers the phase transition. a.** Swapping scheme used to make hetero-polymers with a controlled number of homolog bead pairs (see Methods). **b.** Normalized gyration radius calculated at equilibrium as a function of pairs of same bead type. Examples of 3D snapshots are shown.

**Binding sites configurations is another control parameter: phase diagram.** Next, we systematically studied the equilibrium thermodynamic phases of hetero-polymer chains in several conditions, by simultaneously varying binding affinity $E_{int}$ and the frequency of bead pairs of same type along the polymer (see Methods). In this way, we obtain the phase diagram of the system (**Fig. 5**). Interestingly, we identify a range of affinities $E_{int}$ where the previously discussed configurational effect is observed. Our results show that lower the binding affinities require higher pair frequency to allow the phase transition; conversely, if we increase the affinity, the globular state is achieved at lower frequencies (**Fig. 5**). In agreement with the SBS model ([13]) and general Polymer Physics ([24]), we observe that for sufficiently high values of the binding affinity (here $E_{int} \geq 5.1 k_BT$) the polymer undergoes the phase transition, regardless the pair frequency. On the other hand, at low binding affinities the loops are not stable, the transition does not occur (here $E_{int} \leq 4.1 k_BT$) and therefore the hetero-polymer is an open SAW conformation, for any pair frequency. The thermodynamic phases of the alternating chain corresponding to two different binding affinities (here 4.8 and 5.1 $k_BT$), are a coil and a globular state, respectively. As seen above, the number of types of binding sites is a control parameter of the phase transition. At fixed binder concentration, when the complexity of the chain increases, a greater binding affinity is required for the globular state to be achieved. It is important to stress that we considered rearrangements only involving pairs of consecutive beads of the same type. Of course, other rearrangement schemes (involving e.g. consecutive triplets, quadruplets and so on) can be implemented and likely expand the phase diagram with more articulated parameters settings. Overall, our results suggest that the spatial arrangement of binding site types along the polymer is an important feature to control the coil-globule phase transition.

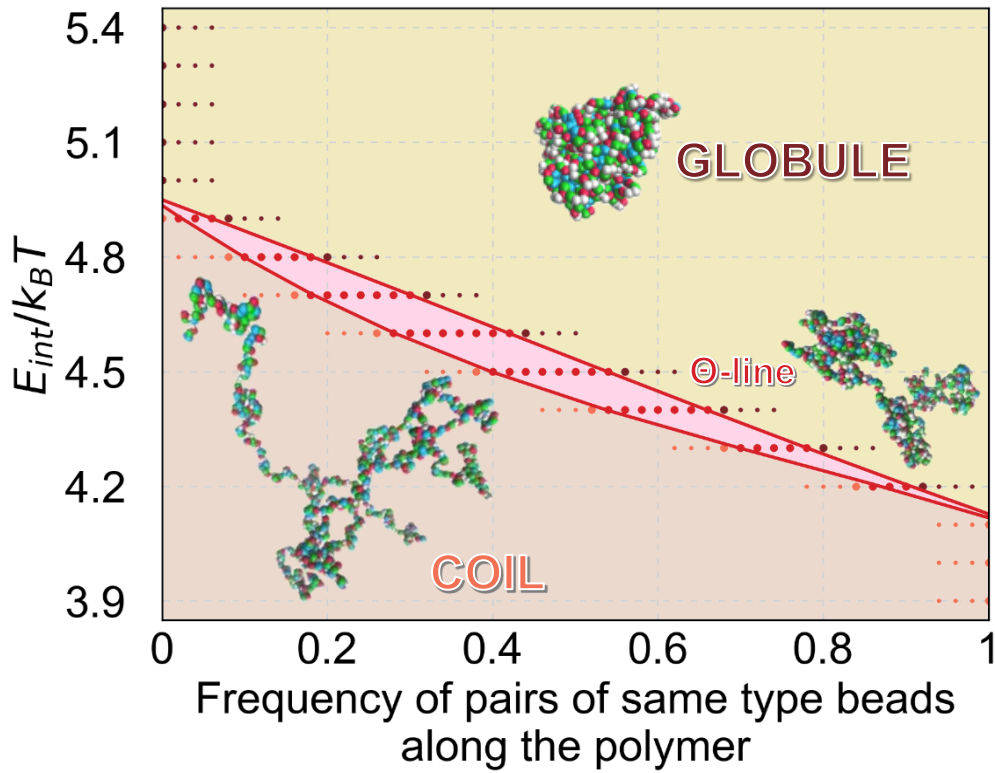

**Fig. 5 Phase diagram of the polymer equilibrium state for different binding affinities and pair frequency.** At low binding affinities or pair frequencies, the polymer is in the open conformation (thermodynamic coil phase). Conversely, for values of these parameters above the threshold, the hetero-polymer is able to reach the globular state. Examples of 3D snapshot for each equilibrium state is also reported.

## DISCUSSION

In this paper, we investigated the coil globule-phase transition of polymer models typically used to describe chromatin architecture ([25]). In particular, we focused on the relationship between the complexity of a hetero-polymer chain and its ability to transit in a compact, globular structure. We found that by increasing the number of molecular types composing the polymer, the system can escape from its stable globular thermodynamic phase. Furthermore, we found that in more complex polymers the transition depends on the arrangement of the binding sites. The phase diagram point toward a scenario in which such mechanism could be a general property of this kind of systems and does not depend on specific parameter settings. Biologically, these results can be linked to the different proteins that typically populate the nuclear environment and that bind to the chromatin filament ([26] Methods) to drive chromatin architecture and genome function. The presence of multiple, different molecular factors that bind to certain genomic regions along with the location of the regulatory elements can be therefore explained as a strategy adopted by the cell to allow a more articulated and precise control of gene activity.

## METHODS

**1. The SBS model.**

According to SBS model ([13]), the polymer chain is modelled as a sequence of N beads and the interactions between distant sites are mediated by binders, modelled as simple, spherical particles. Beads that interact with the binders are named binding sites. At the beginning of the simulation, the polymer is initialized as a SAW (self-avoiding walk) configuration and the binders are randomly placed in the simulation box. Then interaction between the binding sites and cognate binders can induce the formation of loops. The stability of these loops depends on the control parameters of the model, i.e., the binder concentration (c) and the binding affinity ($E_{int}$). If the concentration or energy is above threshold, the loops become stable, and the polymer transits from an open conformation to a globular, compact phase. The thermodynamic phase intermediate to the coil and globule states is known as Θ-point and the corresponding folding state is described by the random walk (RW) model. Binding sites can be specific, i.e. they can interact only with associated cognate binders. Schematically, different types can be identified with different colors. In general, each string is composed of several types of binding sites. For sake of simplicity, in this work we consider the same binding affinity for each color. We consider also an inert, non-interacting type (colored in grey) of bead, which therefore has no cognate binders.

## 2. Details of the Molecular Dynamics (MD) simulations.

Beads and binders are subject to Brownian motion, so each particle obeys a Langevin equation ([27]), solved numerically with a standard velocity Verlet's algorithm using the freely available LAMMPS package ([28]). Interactions are modelled as previously described ([14]). Briefly, all particles experience a repulsive interaction modelled as a truncated, repulsive Lennard-Jones (LJ) potential ([29]), with diameter σ = 1, mass m = 1, energy scale ε = 1 $k_B$T ($k_B$ is the Boltzmann constant and T is the temperature of the system) and distance cut-off $r_{cut}$ = 1.12 σ (we adopted dimensionless units and used the notation given in [29]). Interactions between bead and binders are modelled as a truncated, attractive LJ potential ([14]) with distance cut-off $r_{cut}$ = 1.3 σ and ε depending on the specific parameter setting. Consecutive beads along the polymer are connected by standard FENE springs ([29]), with distance cut-off $R_{int}$ = 1.6 and constant K=30$k_B$T/σ$^2$.

In this work, all hetero-polymer chains are composed of N=1000 beads and the system is confined to a cubic box with periodic boundary conditions and linear size D=70. We explored an affinity (given by the minimum of the LJ attractive potential ([14]) range in the weak biochemical energy scale, in range approximately $E_{int}$ = 3÷5 $k_B$T. For sake of simplicity, we used the same binder concentrations for all the types and considered 500 binders for each color. This value is the same for all the simulations performed. We let the system evolve until stationarity is reached, as shown by the gyration radius as a function of MD iteration steps. We consider an integration timestep $\Delta t$ = 0.012 ([27], [28]) and used 3*10$^7$ timesteps for each MD simulation. For each parameter setting we simulated up to 30 independent runs and the physical quantities of interest ($R_g$/$R_g$(SAW), $P_c$(s) and $R^2$(s), see next sections) are calculated as ensemble averages.

## 3. Types of hetero-polymer chains.

### Alternating and random hetero-polymers.

Suppose m is the number of bead types in a polymer of length N, we design an alternating hetero-polymer in the following way. We take m beads of different types and insert them into the first m sites of the string. From the (m+1)-th position we insert other m beads in the same order as the previous sub-sequence and iterate this mechanism until position N is occupied. In this way, there are no regions with consecutive beads of the same type. Random hetero-polymer chains are generated from alternating ones by randomly permuting the positions of each bead along the string.

### Beads swap and frequency of couples of same type beads along the polymer.

To consider intermediate configurations between the alternate and the randomly arranged polymer previously defines, we designed a procedure that transforms alternating hetero-polymer chains into polymers where the frequency of pairs of similar beads is fixed. The was applied on the polymer having the inert type (grey) and three types of binding sites (red-green-blue in the figures). Specifically, the algorithm takes as input a hetero-polymer chain, where the main sub-sequence is the quadruple containing the colours grey, red, green and blue, grey being the non-interacting particle. To make consecutive pairs of beads of the same type along the polymer, we first select the positions of two differently coloured beads and then swap their position. We start with the grey and blue beads, as schematically shown in **Fig. 3**, **a**. Because of such a substitution, there are now two pairs of same type beads, one blue and one grey. We define $n_{14}$ the number of substitutions. The maximum number of substitutions is 125, since the polymer contains 250 beads of each of the four types. To make polymers with symmetric arrangements of pairs, for a fixed number of substitutions $n^*_{14} \in \{1,...,125\}$, we divide the length of the polymer into $n^*_{14}$ blocks, consider the first quadruple of each block, and swap the grey bead with the blue one within each quadruple. Once made all possible pairs of grey and blue beads, we start again the procedure and swap the beads with unchanged spatial distribution, i.e. green and red types. The number of substitutions is defined $n_{23}$ and again $n^*_{23}=125$ are possible. We define the general parameter $n=n_{14}+n_{23}$ as the number of substitutions of the beads, regardless of their type. By definition $n \in \{0,...,250\}$, where $n=0$ is the alternating polymer with four types of beads. By normalizing this number in the range [0,1], we obtain the pair frequency which is used as control variable in **Figs. 4 and 5**.

**4. Quantities of interest in polymer physics.**

**$R_g/R_g$(SAW) and 3D graphic visualization**.
The gyration radius $R_g$ measures the size of the sphere containing the polymer beads and has been calculated according to its definition: $R_g^2 = \frac{1}{2N^2}\sum_{i,j}^{N}|\bar{R}_i - \bar{R}_j|^2$, being $\bar{R}_i$ the position of i-th bead. To monitor the phase-transition, we used the dimensionless quantity $R_g/R_g$(SAW), i.e. the gyration radius normalized with respect its initial value in the SAW state. 3D structures were generated using the software Visual Molecular Dynamics ([30]).

**Mean contact probability $P_c(s)$ and mean square distance $R^2(s)$**.
The contact probability $P_c(s)$ and quadratic distance $R^2(s)$ were calculated as follows. We define genomic distance s between beads of sites i and j of the polymer as the number of beads separating them, i.e. s=|i-j|. For each s, the mean square distance $R^2(s)$ are computed as the arithmetic average of all possible bead pairs having the same genomic distance s. We then define a distance threshold λσ (we set λ=3.5). If two beads have a distance lower than the threshold they are considered in spatial contact. Mean contact probability is the average frequency of finding two beads at distance s in contact.